\pdfoutput=1 

\documentclass[12pt,preprint]{aastex}

\usepackage{natbib}

\newcommand{\figref}[1]{Figure~\ref{#1}}
\newcommand{\eqref}[1]{equation~(\ref{#1})}

\newcommand{\Alfven}{Alfv\'{e}n\ }

\newcommand{\wpe}{\omega_{\rm{pe}}}
\newcommand{\wce}{\Omega_{\rm{ce}}}

\newcommand{\wci}{\Omega_{\rm{ci}}}
\newcommand{\Ma}{M_{\rm{A}}}
\newcommand{\Va}{v_{\rm{A}}}

\newcommand{\mug}{\mu \rm{G}}

\shorttitle{Electron Surfing Acceleration in Multidimensions}
\shortauthors{Amano \& Hoshino}

\begin{document}
\title{Electron Shock Surfing Acceleration in Multidimensions: Two-dimensional Particle-In-Cell Simulation of Collisionless Perpendicular Shock}
\author{\sc Takanobu Amano\altaffilmark{1} and Masahiro Hoshino}
\email{amanot@stelab.nagoya-u.ac.jp}
\affil{Department of Earth and Planetary Science, University of Tokyo, 7-3-1 Hongo, Bunkyo-ku, Tokyo 113-0033, Japan}
\altaffiltext{1}{Now at Solar-Terrestrial Environment Laboratory, Nagoya University, Furo-cho, Chikusa-ku, Nagoya 464-8601, Japan}

\begin{abstract}
Electron acceleration mechanism at high Mach number collisionless shocks propagating in a weakly magnetized medium is investigated by a self-consistent two-dimensional particle-in-cell simulation. Simulation results show that strong electrostatic waves are excited via the electron-ion electrostatic two-stream instability at the leading edge of the shock transition region as in the case of earlier one-dimensional simulations. We observe strong electron acceleration that is associated with the turbulent electrostatic waves in the shock transition region. The electron energy spectrum in the shock transition region exhibits a clear power-law distribution with spectral index of $2.0 {\rm -} 2.5$. By analyzing the trajectories of accelerated electrons, we find that the acceleration mechanism is very similar to shock surfing acceleration of ions. In contrast to the ion shock surfing, however, the energetic electrons are reflected by electron-scale electrostatic fluctuations in the shock transition region, but not by the ion-scale cross-shock electrostatic potential. The reflected electrons are then accelerated by the convective electric field in front of the shock. We conclude that the multidimensional effects as well as the self-consistent shock structure are essential for the strong electron acceleration at high Mach number shocks.
\end{abstract}

\keywords{acceleration of particles --- cosmic rays --- plasmas --- shock waves}

\section{INTRODUCTION}
It is generally believed that cosmic rays with energies up to the knee ($\sim 10^{15}$ eV) are produced at supernova remnant (SNR) shocks. There is indeed direct evidence for shock acceleration of cosmic ray electrons to more than TeV energies \citep[e.g.,][]{1995Natur.378..255K}. Recently, TeV gamma-rays from some shell-type SNRs have been detected by HESS, which implies the presence of cosmic rays with $\sim 100$ TeV energies \citep{2007A&A...464..235A}. Although it is still under active debate whether the primary particles emitting the gamma-rays are either electrons or protons, the morphological similarity between nonthermal X-ray emission and the gamma-rays indicates that they are accelerated by SNR shocks. Diffusive shock acceleration (DSA) is the most widely accepted theory for the shock acceleration of nonthermal particles \citep[e.g.,][]{1987PhR...154....1B}. The DSA theory assumes the presence of magnetohydrodynamic (MHD) turbulence upstream of the shock. Energetic particles scattered by MHD waves gain energy by diffusively crossing the shock front back and forth many times. The central unresolved issue in DSA theory is the well-known injection problem: Since DSA is efficient only for particles having energy enough to be scattered by MHD waves, injection from a thermal pool to nonthermal energies by some other mechanism is required. This requirement is very stringent particularly for electrons because of their small Larmor radii. Therefore, strong preacceleration mechanics are needed to explain the observed nonthermal emissions from ultra-relativistic electrons at SNRs. Numerical studies using particle-in-cell (PIC) codes have been conducted to explore the possibilities of direct electron energization at the vicinity of the shock that may provide a seed population for DSA \citep[e.g.,][]{2000A&A...356..377D,2000ApJ...543L..67S,2001PhRvL..87y5002M,2002ApJ...572..880H,2002ApJ...579..327S}. \cite{2007ApJ...661..190A} have recently shown that a fraction of electrons may efficiently be injected to DSA process at high Mach number quasi-perpendicular shocks. Their one-dimensional (1D) PIC simulations demonstrated that nonthermal electrons are generated by successive two different acceleration mechanisms, namely, shock surfing acceleration (SSA) and shock drift acceleration (SDA). They proposed an electron injection model based on the 1D simulation results, which can account for the observed injection efficiencies \citep[e.g.,][]{2003ApJ...589..827B}. However, the problem is that comprehensive theory of SSA does not exist at present. Therefore, the effects of multidimensionality on the injection efficiency were not taken into account. Two- or three-dimensional self-consistent numerical simulations of high Mach number shocks are needed to evaluate the realistic injection efficiency.

It is well-known that an important portion of dissipation at collisionless non-relativistic shock is provided by the so-called reflected ions. At quasi-perpendicular shock with $\theta_{\rm Bn} > 45 \degr$ ($\theta_{\rm Bn}$ is an angle between the shock normal and the upstream magnetic field), the reflected ions gyrating in front of the shock are accelerated by the convective electric field in the upstream region, and then transmitted to the downstream. Early hybrid simulation (kinetic ions and massless electrons) studies showed that the direct energization of the reflected ions contributes importantly to the downstream thermalization \citep[e.g.,][]{1982JGR....87.5081L}. On the other hand, it is generally considered that the energization of electrons at collisionless shock is relatively weak. Since Larmor radii of electrons are very small compared to the scale length of macroscopic electromagnetic fields, they are considered to suffer only adiabatic heating by the compressed magnetic field at the shock. In contrast, the in-situ observations of the Earth's bow shock demonstrated that this is not always true \citep{1989JGR....9410011G,2006GeoRL..3324104O}. Furthermore, radio and X-ray observations strongly suggest that nonthermal electron acceleration is very efficient at young SNRs. Microinstabilities in the shock transition region probably play an important role for nonadiabatic energization of electrons. Recent PIC simulations of quasi-perpendicular shocks have shown that a variety of instabilities can be excited in the shock transition region by the presence of the reflected ions \citep[e.g.,][]{2000ApJ...543L..67S,2004PhPl...11.1840S,2003JGRA..108.1014S,2003JGRA..108.1459M,2006JGRA..11106104M,2006AdSpR..37..483M}. Among them, the Buneman instability \citep{1958PhRvL...1....8B}, which is the electrostatic two-stream instability between cold electrons and ions, is the most dominant mode at high Mach number regime relevant to SNRs. \cite{2002ApJ...572..880H} showed that the Buneman instability plays a key role for the production of nonthermal electrons via SSA. Strong electrostatic potential produced by the nonlinear evolution of the instability can trap a fraction of electrons; the trapped electrons moving with the wave potential can see an inductive electric field arising from the relative velocity between the wave and the background plasma. Therefore, they can be accelerated in the transverse direction until they escape from the potential. Nonlinear 1D PIC simulations demonstrated that SSA can quickly accelerate electrons to mildly relativistic energies \citep{2001PhRvL..87y5002M,2002ApJ...572..880H}. Therefore, it is believed that SSA plays an important role for an efficient electron injection. It is worth noting that the electron energization by SSA relies on the assumption that the potential is uniform in the transverse direction, so that the electron transport in the direction along the inductive electric field is very efficient. However, it is well-known that the Buneman instability at oblique propagation has growth rates comparable to the parallel propagation \citep{1974PhFl...17.428}. The assumption of one-dimensional wave potential is, therefore, not appropriate to evaluate the realistic efficiency of SSA. \cite{2007ApJ...661L.171O} recently pointed out by performing two-dimensional (2D) electrostatic PIC simulations that SSA may be inefficient in multidimensions. Their conclusion was drawn from the observation that they did not observe nonthermal tails in the final electron energy spectra. However, they used a homogeneous model of the shock transition region in which the plasma was consisting of three components, the upstream electrons and ions, and the reflected ion beam. We think the artifact introduced by their model should be taken into account with great care.

Here we report 2D PIC simulation results of a high Mach number, perpendicular shock propagating in a weakly magnetized plasma. Note that several numerical studies of collisionless shocks using 2D PIC codes can be found in the literature \cite[e.g.,][]{1984JGR....89.2142F,1992PhFlB...4.3533L}. However, these studies considered only moderate Mach number shocks relevant to the Earth's bow shock. At higher Mach number regime, we find that strong electrostatic waves in the shock transition region excited by the Buneman instability as in the case of 1D simulations. Efficient electron acceleration associated with the large amplitude electrostatic waves is observed. It is shown that the nonthermal electrons are produced by a mechanism similar to SSA of ions \citep[e.g.,][]{1996JGR...101..457Z,1996JGR...101.4777L}. We argue that the effects of multidimensionality and the self-consistent shock structure are essential for the production of nonthermal electrons at high Mach number shocks.

\section{SIMULATION}
\subsection{Simulation Setup}
We use a 2D electromagnetic PIC simulation code, in which both electrons and ions are treated as kinetic macroparticles, to study the dynamics of electrons and ions in a fully self-consistent shock structure. A shock wave is excited by the so-called injection method, that is commonly used in 1D simulations. A high-speed plasma consisting of electrons and ions is injected from the boundary $x = 0$ of a 2D simulation box in the $x {\rm -} y$ plane and travels toward the positive $x$ direction. The plasma carries the uniform magnetic field perpendicular to the simulation box (${\bf B}_0 \parallel {\bf e}_z$). At the opposite boundary, particles are specularly reflected by the wall. Then, a perpendicular shock forms through the interactions between the incoming and the reflected particles, and it propagates in the negative $x$ direction. Therefore, the simulation is done in the downstream rest frame. The periodic boundary condition is imposed in the $y$ direction.

We use the following plasma parameters in the upstream: $\beta_e = \beta_i = 0.5$ ($\beta_j \equiv 8 \pi n T_j / B^2$), where $n$, $T_j$ and $B$ are the density, temperature, and magnetic field strength, respectively. The ratio of the plasma frequency to the electron cyclotron frequency is $\wpe/\wce = 10$. A reduced ion to electron mass ratio of $m_i/m_e = 25$ is used. These lead to a upstream \Alfven speed of $\Va/c = 0.02$. We use a plasma injection four-velocity of $U_0/c = 0.2$. The \Alfven Mach number of the resulting shock wave is $\Ma \simeq 14$ in the shock rest frame. The grid size of the simulation is taken to be equal to the electron Debye length in the upstream. We use $4096 \times 256$ grid points in the $x$ and $y$ direction, respectively. The physical size of the simulation box is $L_x \simeq 204\,c/\wpe$ and $L_y \simeq 12.8\,c/\wpe$. Initially, each cell contains 40 particles for each species in the upstream. Note that the injection velocity of $U_0/c = 0.2$ is rather high for simulations of realistic SNR shocks. We adopt this value to reduce the computational costs. However, the dominant instability in the present simulation is still the electrostatic mode in contrast to an electromagnetic Weibel-like instability found in relativistic shocks \citep[e.g.,][]{2007ApJ...668..974K}. Hence, we think the essential physics does not change due to the use of an artificially high shock speed.

We use the following units unless otherwise stated: time, distance, velocity, energy will be given in units of the inverse of the electron plasma frequency in the upstream $\wpe^{-1}$, the electron inertial length $c/\wpe$, the injection velocity $U_0$, the upstream electron bulk energy $\epsilon_0 = (\gamma_0 - 1) m_e c^2$ where $\gamma_0 = \sqrt{1 + (U_0/c)^2}$, respectively. The electric and magnetic field are normalized to the motional electric field $E_0 = U_0 B_0 / \gamma_0 c$, and the background magnetic field $B_0$ in the upstream.

\subsection{Shock Structure}
\begin{figure}[t]
 \figurenum{1}
 \epsscale{0.75}
 \plotone{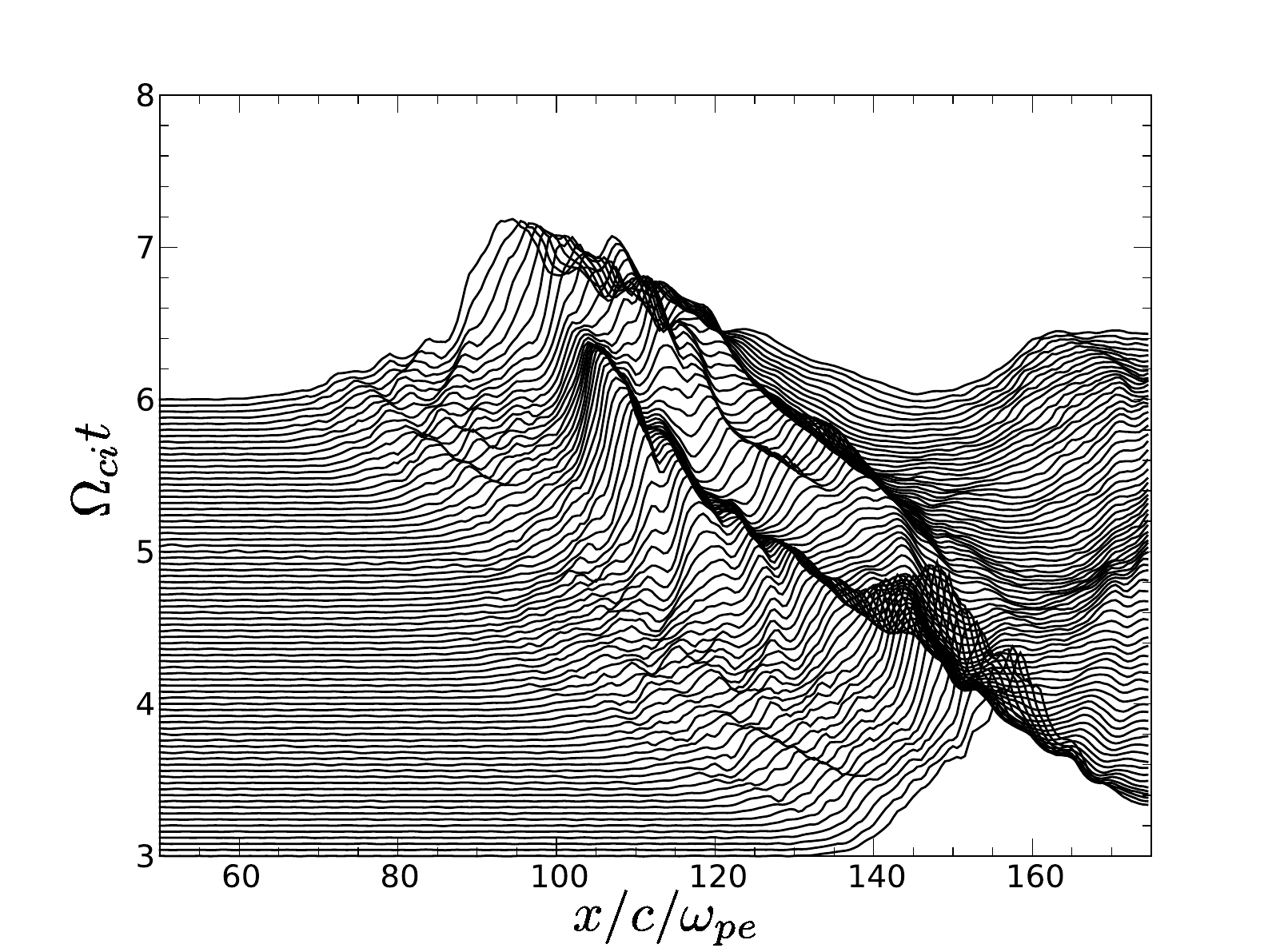}

 \caption[Stacked profiles of magnetic field]{Stacked profiles of compressional magnetic field component $B_z$, averaged over the $y$ direction. The vertical axis is normalized to the inverse of ion cyclotron frequency in the upstream.}

 \label{stackplot}
\end{figure}

We first discuss an overall structure of the simulated high Mach number shock. \figref{stackplot} shows the stacked magnetic field profiles $B_z$ averaged over the $y$ direction. In this figure, we can clearly see a shock wave propagating in the negative $x$ direction. The average shock propagation speed is about $\sim 0.4 U_0$, yielding a Mach number of $\Ma \sim 14$ in the shock rest frame. Note that the vertical axis is normalized to the inverse of ion cyclotron frequency in the upstream $\wci^{-1}$ ($\wci t = 250 \; \wpe t$). The shock propagation is not stationary, but shows slight variation in the shock structure. It is known that quasi-perpendicular shocks with high Mach numbers simulated by 1D PIC codes typically show nonstationary behavior called cyclic self-reformation, which occurs on a characteristic timescale of $1\rm{-}2$ $\wci^{-1}$. However, the observed shock front shows less time variability than usually observed in 1D, suggesting that an efficient plasma thermalization is suppressing the nonstationary behavior \citep{2004AnGeo..22.2345S,2005JGRA..11002105S}.

\begin{figure}[t]
 \figurenum{2}
 \epsscale{0.95}
 \plotone{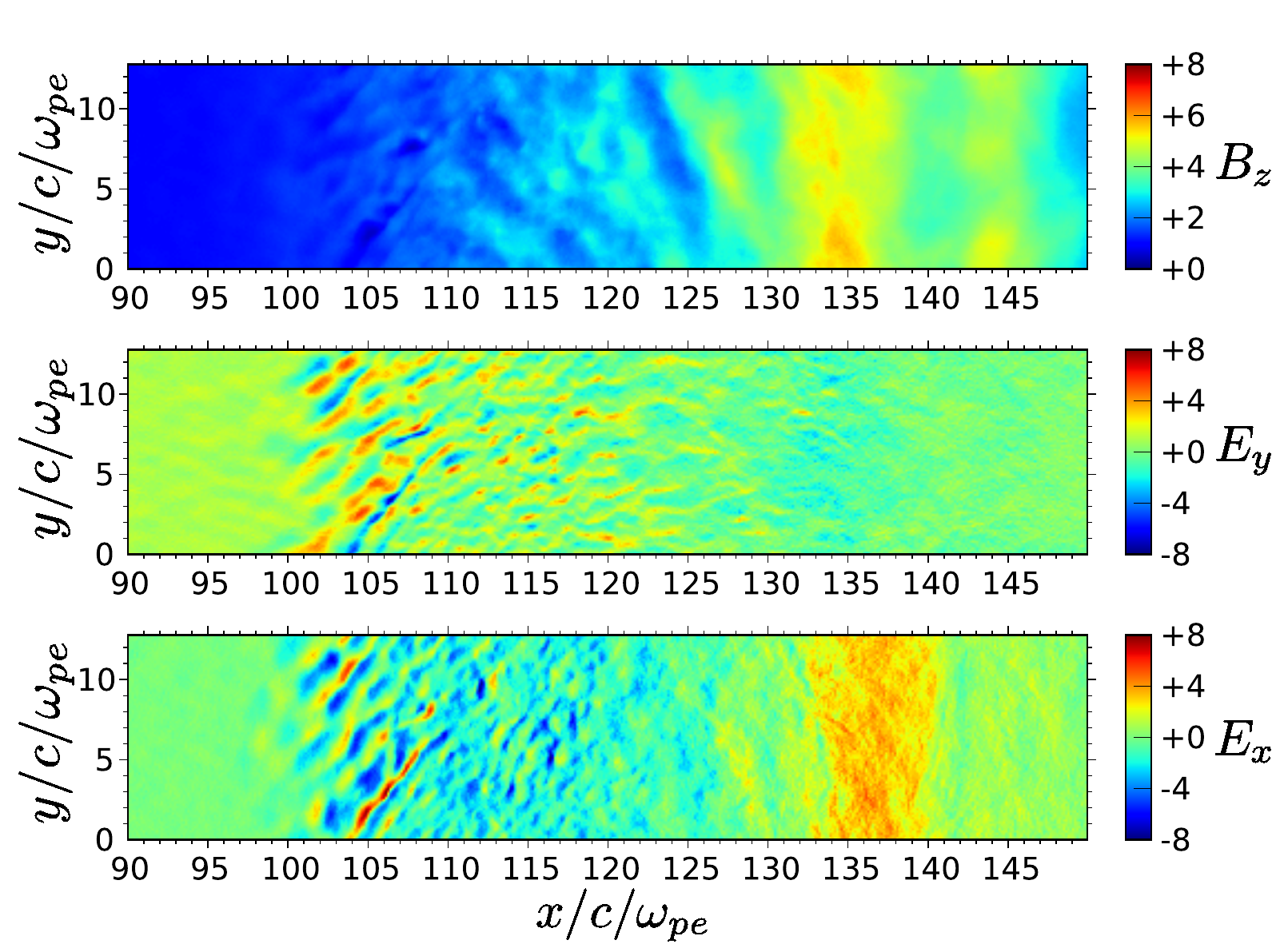}

 \caption[Snapshot of electromagnetic field]{Snapshot of electromagnetic fields at $\wpe t = 1000$.  From top to bottom, color images of $B_z$, $E_y$, $E_x$ are shown.}

 \label{field-1000}
\end{figure}

Shown in \figref{field-1000} is the snapshot of the electric and magnetic field at $\wpe t = 1000$. The leading edge of the shock transition region is located at around $x / c/\wpe \simeq 100$. We can see predominantly electrostatic fluctuations at $100 \lesssim x / c/\wpe \lesssim 110$ in both $E_x$ and $E_y$ panels. These waves are excited via the Buneman instability caused by the interactions between the upstream electrons and the reflected ions. It should be noted that the wavefronts of these electrostatic waves are oblique to the shock normal, which is in sharp contrast to 1D. Furthermore, the waves are not one-dimensional, having finite extent along the wavefront. The excitation of multidimensional wave structure by the Buneman instability is consistent with the linear theory and nonlinear 2D PIC simulations in a periodic simulation box \citep{1974PhFl...17.428}.

\begin{figure}[p]
 \figurenum{3}
 \epsscale{0.95}
 \plotone{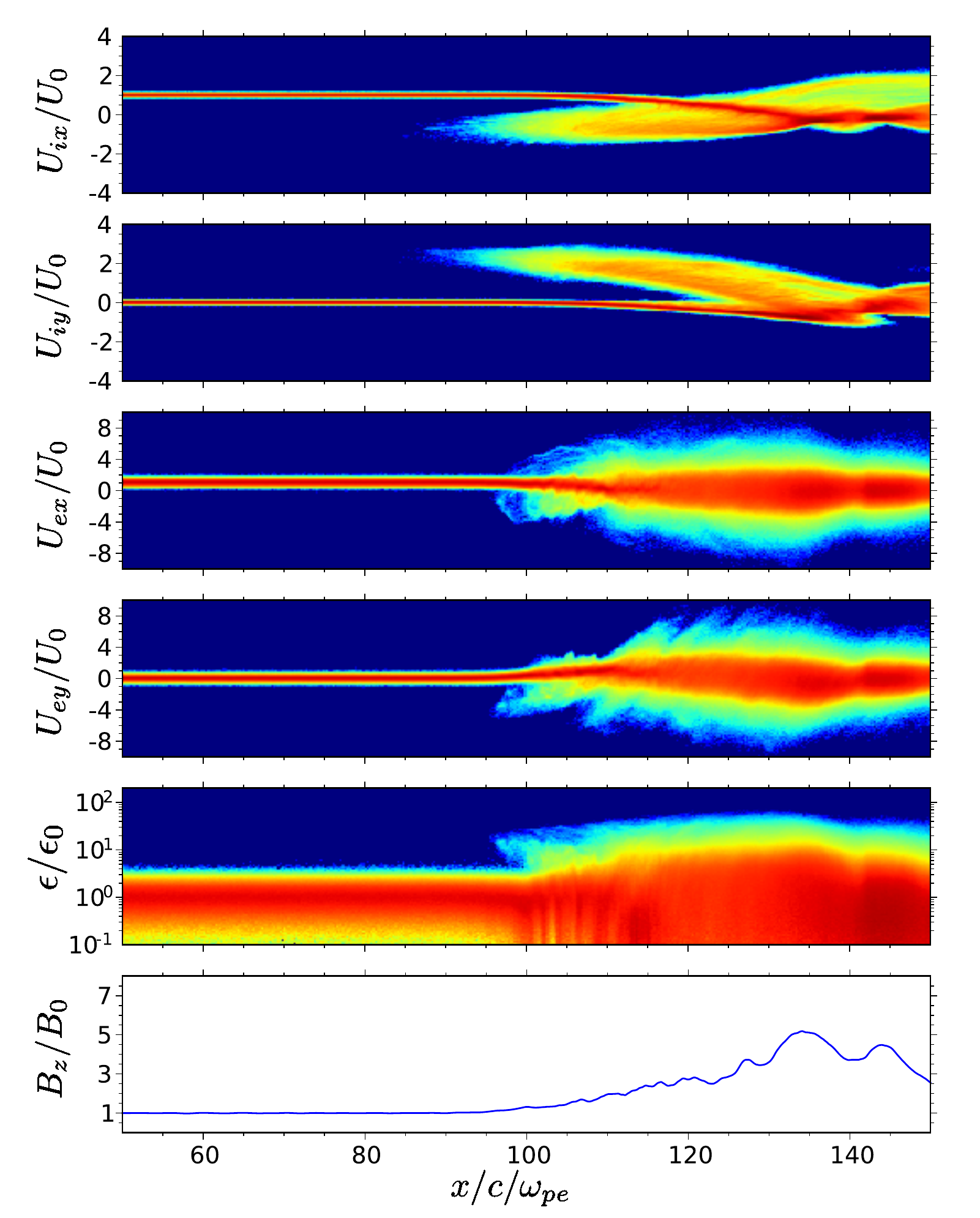}

 \caption[Snapshot of particle phase-space plot]{Snapshot of particle phase-space plots and compressional magnetic field profile averaged over the $y$ direction. Color represent the logarithm of the particle count in each bin. Note that the vertical scale of the electron energy spectrum (the second panel from the bottom) is shown on a logarithmic scale.}

 \label{overview-1000}
\end{figure}

The reason why we observe the oblique wavefronts can easily be understood by considering the Larmor motion of the reflected ions. \figref{overview-1000} displays the snapshot of the phase-space of both electrons and ions, as well as the $y$-averaged magnetic field. The reflected ions can easily be identified in the top two panels showing the phase-space plots of ions in $(x, u_{i,x})$ and $(x, u_{i,y})$. Since the reflected ions are accelerated in the positive $y$ direction, they have a large bulk velocity not only in the $x$, but also $y$ direction at the leading edge of the shock $x/c/\wpe \sim 100$. The waves excited by the Buneman instability propagate mostly parallel to the beam direction. Therefore, it is not surprising that we observe the oblique wavefronts. To be more precise, the instability excites a wide range of oblique modes and the observed spatial profile (wavefront) is a superposition of waves with different wave vectors. However, we observe the oblique wavefronts propagating parallel to the beam probably because (1) the wave power peaks at the parallel propagation, and (2) the wave propagation is symmetric with respect to the beam. We have actually confirmed that the superposed spatial profile propagates almost perpendicular to the wavefronts (parallel to the beam). It is worth noting that this behavior agrees very well with that observed in periodic simulations of the Buneman instability in 2D.

Looking at the electron phase-space plots $(x, u_{e,x})$, $(x, u_{e,y})$, that are shown below the ions, we can find strong electron energization at the leading edge of the shock transition region $x/c/\wpe \sim 100$. It is clear that the energization of electrons is associated with the strong electrostatic waves excited by the Buneman instability as had been studied by 1D codes. As we see below, however, the energization of electrons in 2D occurs in a somewhat different manner, which is due to the different properties of the strong electrostatic turbulence in the foot region.

\subsection{Energy Spectrum}
\begin{figure}[t]
 \figurenum{4}
 \epsscale{0.75}
 \plotone{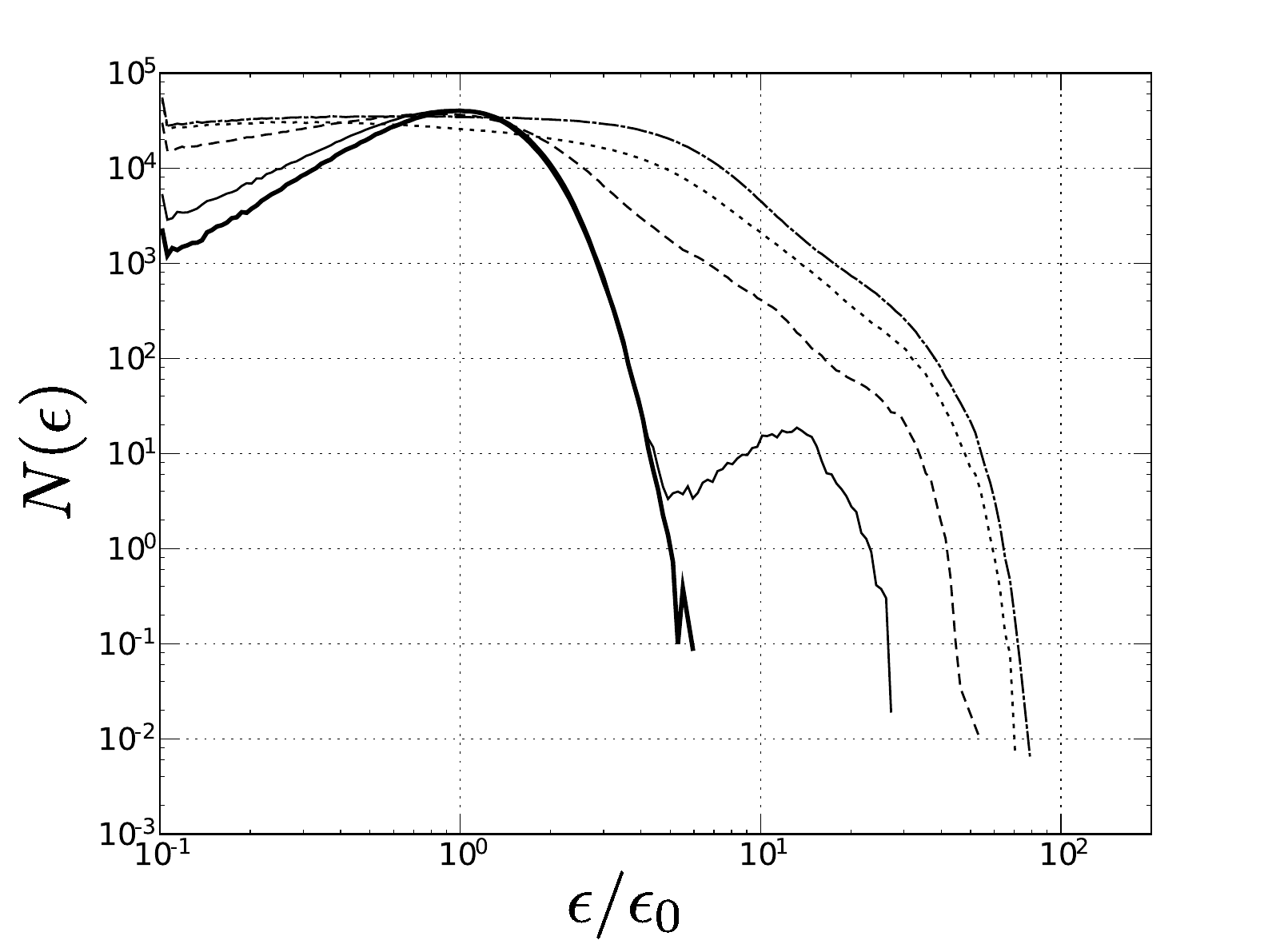}

 \caption[Energy spectra of electrons]{Energy spectra of electrons around the shock transition region at $\wpe t = 1000$. Each line shows a spectrum averaged over $75 \le x/c/\wpe \le 87.5$ (thick solid), $87.5 \le x/c/\wpe \le 100$ (solid), $100 \le x/c/\wpe \le 112.5$ (dashed), $112.5 \le x/c/\wpe \le 125$ (dotted), $125 \le x/c/\wpe \le 137.5$ (dash-dotted), respectively.}

 \label{spectrum-1000}
\end{figure}

The electron energy spectra shown in \figref{overview-1000} are integrated over every $12.5 c/\wpe$ interval to obtain \figref{spectrum-1000} showing the averaged energy spectra around the shock transition region. One can clearly find power-law energy spectra within the shock transition region. The observed spectral slopes are $\sim 2.0 {\rm -} 2.5$. The slope slightly steepens with increasing the penetration into the shock. The downstream spectrum is essentially unchanged from the spectrum observed at the overshoot $125 \le x/c/\wpe \le 137.5$ (dash-dotted line). The steeper spectral indices observed in the deeper shock transition region suggest that the nonthermal electrons are mostly produced at the leading edge of the shock transition region. It is worth noting that a high energy hump is observed in the distribution right before the shock transition region $87.5 \le x/c/\wpe \le 100$ (solid line). This hump corresponds to energetic electrons that are once reflected and are gyrating in front of the shock.

\subsection{Particle Acceleration}
\begin{figure}[t]
 \figurenum{5}
 \epsscale{1.0}
 \plotone{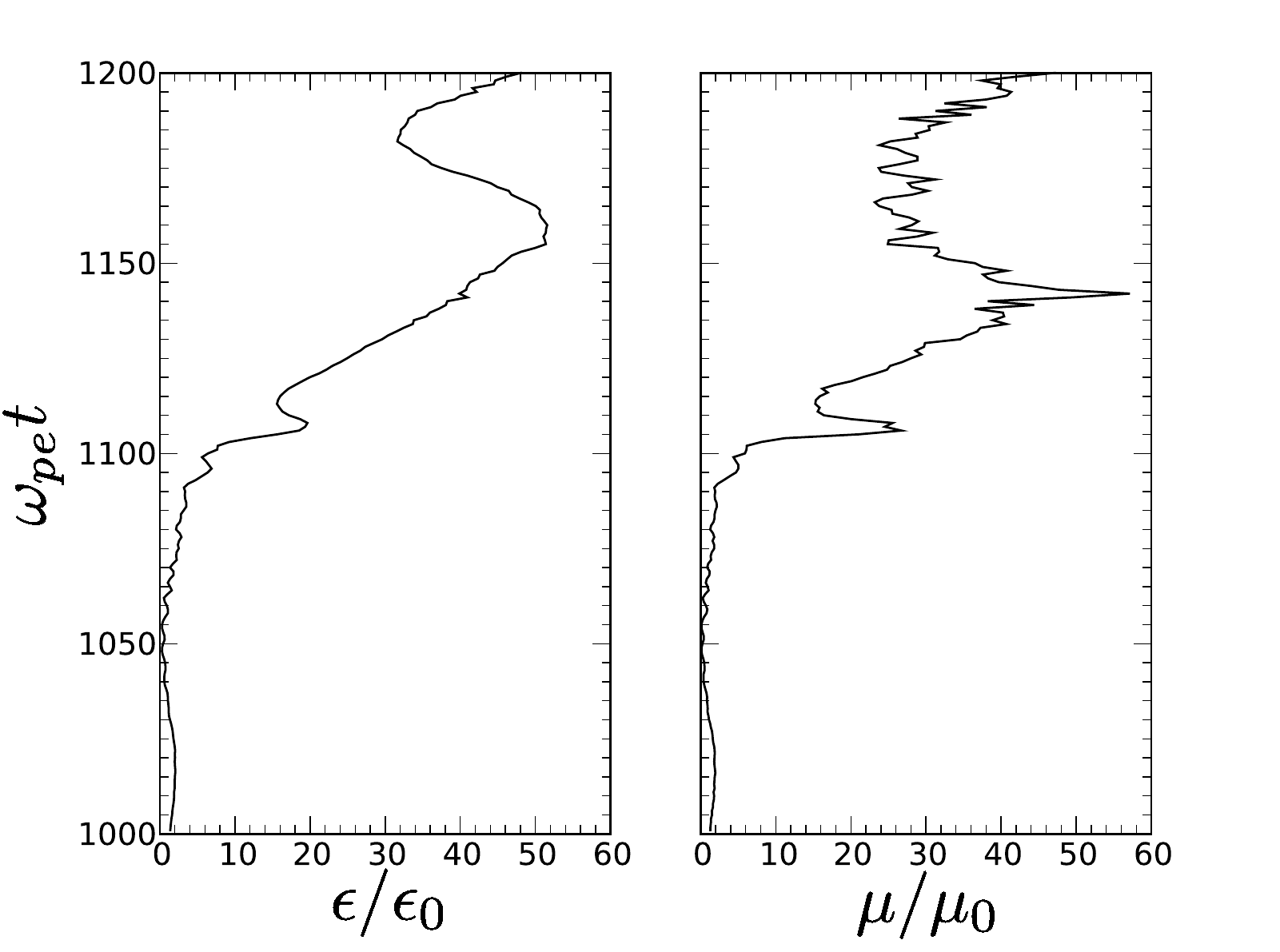}

 \caption[Time history of accelerated electron]{Time history of an accelerated electron: energy (left) and the first adiabatic invariant (right).}

 \label{trajectory-energy}
\end{figure}

\begin{figure}[t]
 \figurenum{6}
 \epsscale{1.0}
 \plotone{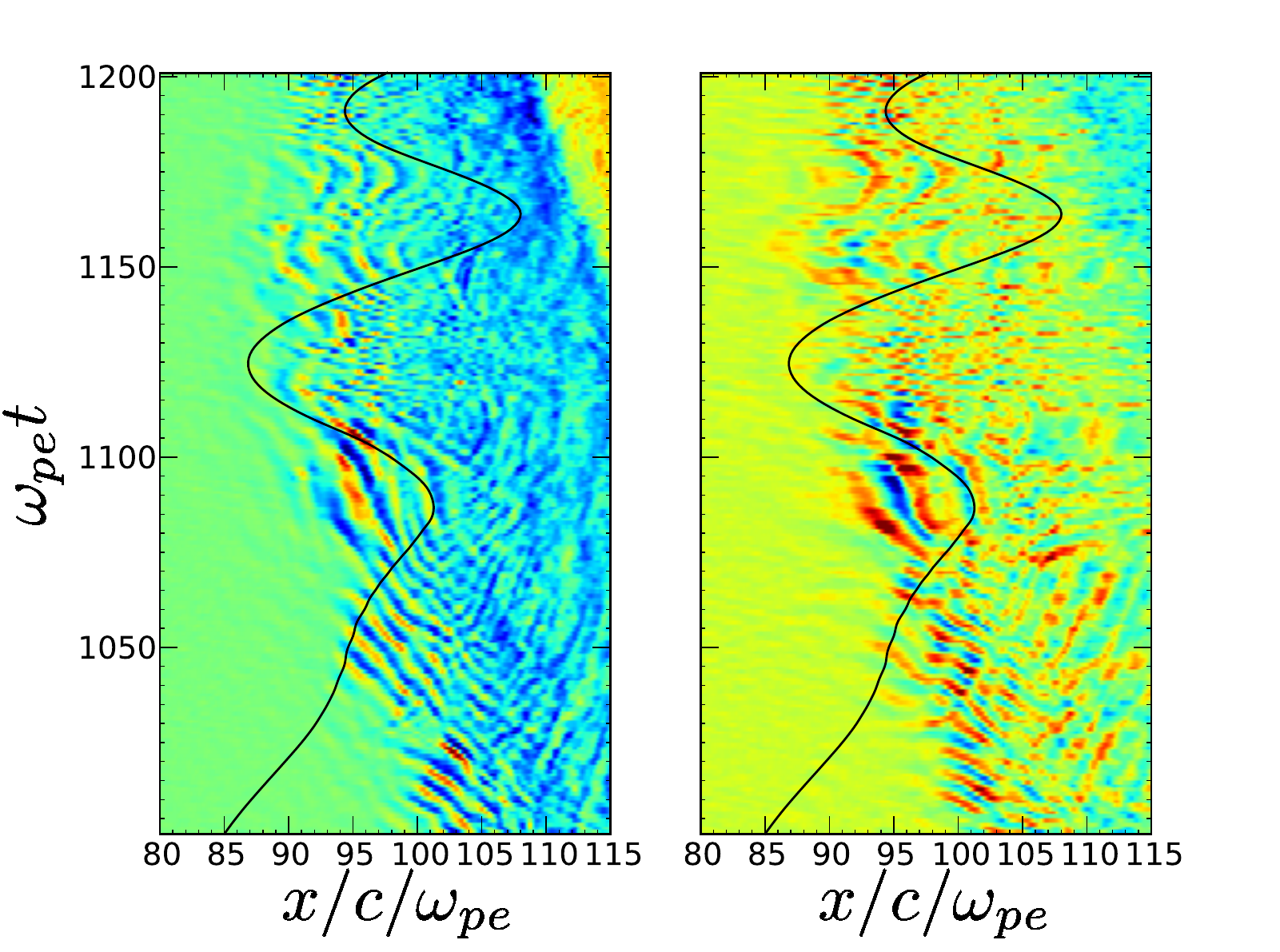}

 \caption[Electron trajectory and electric field]{Electron trajectory and electric fields $E_x$ (left), and $E_y$ (right), respectively.}

 \label{trajectory-exey}
\end{figure}

\begin{figure}[t]
 \figurenum{7}
 \epsscale{1.0}
 \plotone{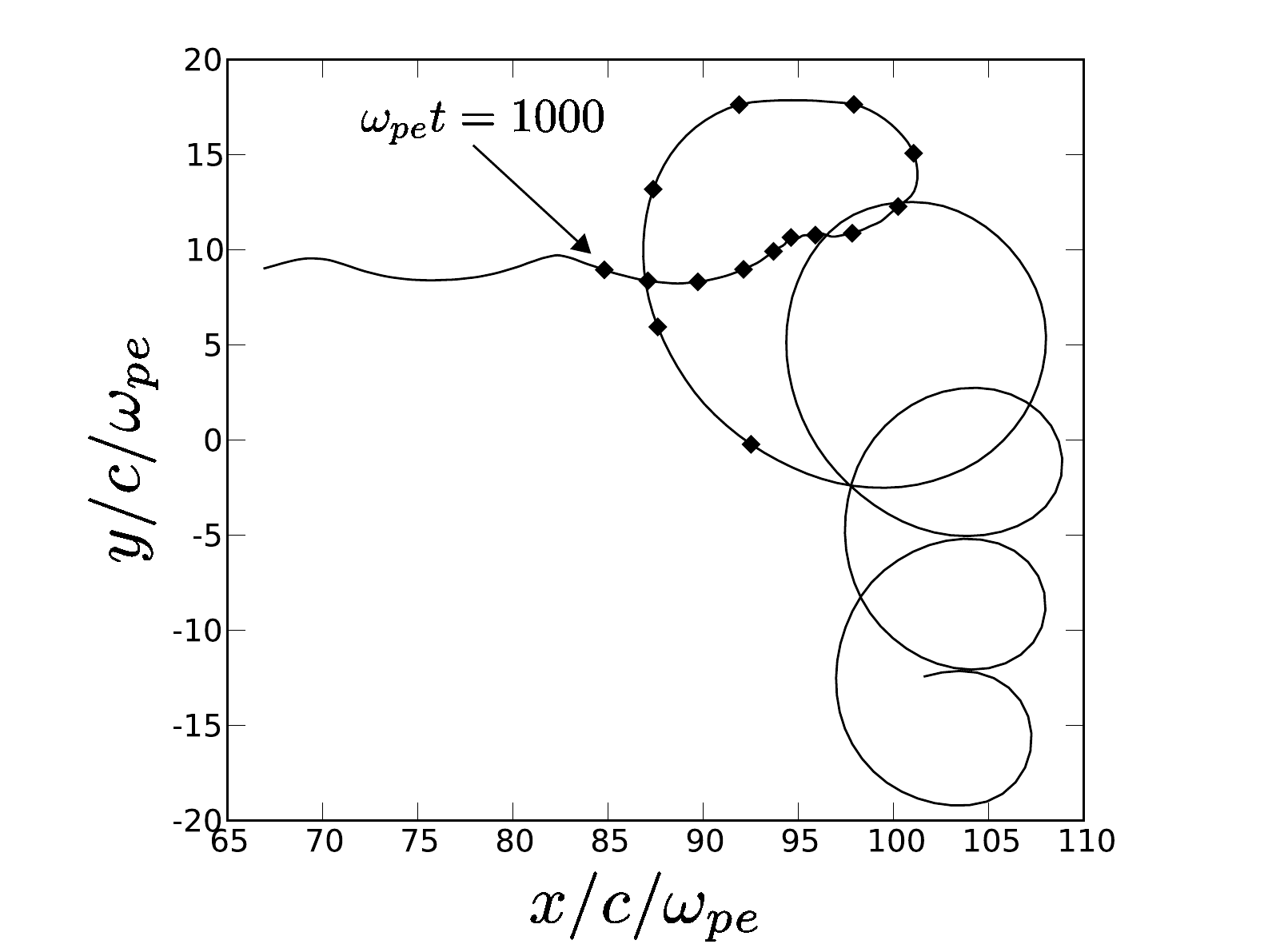}

 \caption[Electron trajectory in the $x{\rm -}y$ plane]{Electron trajectory in the $x{\rm -}y$ plane. Symbols are plotted every $\wpe \Delta t = 10$ interval from $\wpe t = 1000$, during which the electron suffers the strong energization.}

 \label{trajectory-xy}
\end{figure}

In order to discuss particle acceleration mechanism in more detail, individual trajectories of energetic electrons are analyzed. \figref{trajectory-energy} shows the time history of energy and the first adiabatic invariant of a typical accelerated electron. Here the first adiabatic invariant is defined as $\mu \equiv u_{\perp}^2 / 2 B$, and is normalized to its upstream value $\mu_0 = U_0^2 / 2 B_0$. We use the downstream rest frame as a reference frame to define the adiabatic invariant. The particle's position $x$ is also plotted in \figref{trajectory-exey} as a function of time. The color shows the electric field $E_x$ (left) and $E_y$ (right) components, respectively. Note that the electric fields shown in the figure are measured at $y = y_p$ for each time step: $E_{x,y} = E_{x,y}(x, y_p, t)$, where $y_p$ represents the particle's position $y$. Thus, one may consider that the particle sees the time variation of a one-dimensional shock structure. The particle trajectory in the $x {\rm -} y$ plane is shown in \figref{trajectory-xy}. The symbols are plotted every $\wpe \Delta t = 10$ interval during the strong electron energization $\wpe t = 1000 {\rm -} 1140$.

The low-energy electron that is initially located in the far upstream region begins to interact with the shock at $\wpe t \simeq 1040$. It sees large amplitude waves which have both $E_x$ and $E_y$ components during $\wpe t \simeq 1040 {\rm -} 1090$, and is gradually heated. Because of the compressed magnetic field as well as the change of the convective electric field, the guiding center velocity slows down in the shock transition region. Hence, the particle trajectory is strongly deflected at $\wpe t \simeq 1090$. After that, the particle is convected toward the negative $x$ direction and are finally ejected into the upstream at $\wpe t \simeq 1110$. At this time, the electron energy already increases by a factor of $\sim 20$. During the stay in the upstream region, it sees the constant upstream convective electric field. Hence the particle is accelerated in the negative $y$ direction during its half gyration. When it returns back to the shock $\wpe t \simeq 1140$, the energy is increased by a factor of $\sim 40$ from its initial value. During this acceleration phase, the particle's first adiabatic invariant also increases by a factor of $\sim 40$. Note that, since the first adiabatic invariant is defined in the downstream frame (not in the guiding center frame), it oscillates with the electron cyclotron period. In addition, its temporal average should also change as the particle passes through the shock even when the particle motion is strictly adiabatic. However, this change is only of the order of unity in the normalized unit, while the particle's first adiabatic invariant increases more than an order of magnitude. Thus, it is obvious that the acceleration is a nonadiabatic process. After $\wpe t \sim 1150$, the particle energy further increases, but only with adiabatically, due to the compressed magnetic field at the shock.

\begin{figure}[t]
 \figurenum{8}
 \epsscale{0.8}
 \plotone{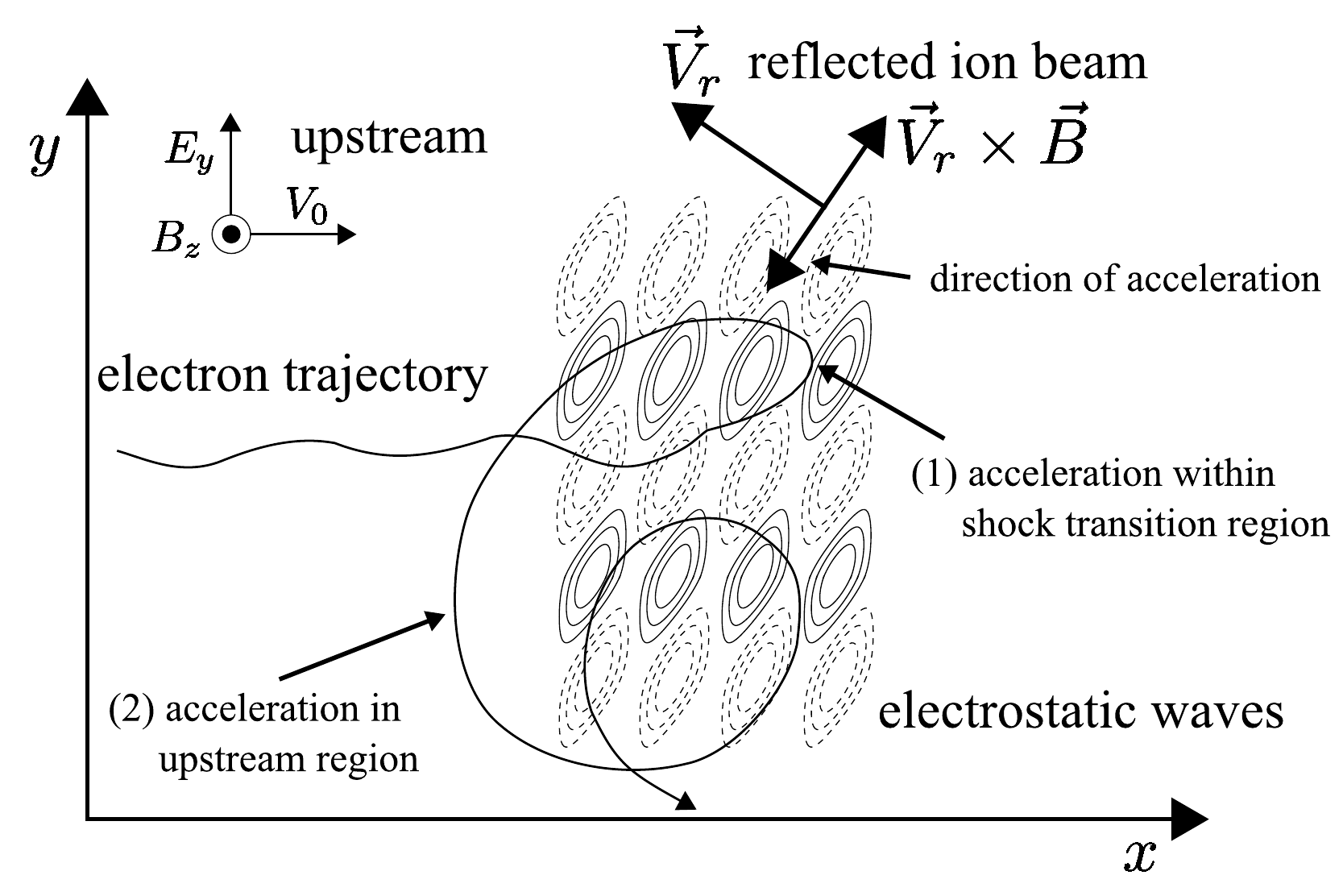}

 \caption[Schematic illustration of acceleration mechanism]{Schematic illustration of acceleration mechanism. Electrons are accelerated in the two steps : (1) they are accelerated by the stochastic electron shock surfing in the shock transition region and preferentially transported to the upstream region. (2) the accelerated electrons escaping into the upstream suffer further accelerated by the constant motional electric field.}

 \label{mechanism}
\end{figure}

The particle acceleration process shown above is considered to be a combination of two mechanisms: one is the energization in the shock transition region, and another is the acceleration in the upstream region (see \figref{mechanism}). We think that the former acceleration mechanism may be understood as a stochastic acceleration by large amplitude electrostatic turbulence. Consider an electron that encounters a large amplitude electrostatic wave. If the electron encounters the wave at a certain gyrophase such that the particle velocity in the direction of the wave propagation is approximately equal to the phase velocity, it can travel (or resonate) with the wave during a certain time interval. Since the wave profile propagates with the speed of the reflected ion beam (which differs from the background plasma flow speed), the resonant particle can see an inductive electric field in the wave rest frame. Therefore, electrons are accelerated in the transverse direction that is parallel to the wavefronts. The mechanism of particle acceleration is similar to SSA in 1D \citep{2001PhRvL..87y5002M,2002ApJ...572..880H}, however, the difference is that the accelerated particles are not trapped in any waves. Instead, they quickly move from one wave to another in a stochastic manner, and are accelerated when they are in resonance with the wave. Here we would emphasize two important characteristics of the electrostatic turbulence: (1) the wavefronts are oblique to the shock normal, and (2) the turbulent region has a finite extent along the shock normal. Since the direction of electron acceleration is approximately anti-parallel to the inductive electric field, the accelerated electrons are preferentially transported in the upstream direction as schematically shown in \figref{mechanism}. Furthermore, since the turbulent region has a finite extent, the accelerated electrons can eventually escape into the upstream of the shock front. It should be noted that the electron reflection is not induced by a macroscopic field, such as the magnetic field gradient, and the cross-shock electrostatic potential: Since the Larmor radii of electrons is very small, the deflection by the magnetic field alone cannot explain the observed reflection. Similarly, the shock potential cannot reflect the negative charge. Indeed, we do not find any reflected electrons in 1D PIC simulations of perpendicular shocks \citep[e.g.,][]{2002ApJ...572..880H}. The strong and multidimensional turbulent electrostatic waves do play a role in the transport of the energetic electrons. We also note that the electron reflection is not an artifact of the use of a small ion to electron mass ratio. Although the scale length of the shock transition region is proportional to the ion Larmor radius, the region of strong electrostatic turbulence always appears at the leading edge of the shock and the scale length of the region depends only weakly on the mass ratio $\propto ( m_i/m_e )^{1/3}$ \citep{1988Ap&SS.144..535P}. Therefore, it is reasonable to expect that the electron reflection occurs at shocks with realistic mass ratios.

The latter acceleration in the upstream can easily be understood by analogy with SSA of ions. The ion shock surfing is caused by the shock potential, that reflects positively charged particles. A fraction of ions reflected by the shock potential can be accelerated by the constant motional electric field during the Larmor motion in the upstream. On the other hand, the accelerated electrons observed here are reflected by the microscopic turbulent electrostatic waves. As a result, they suffer further acceleration by the motional electric field in the upstream. Because of this similarity, we consider that the present electron acceleration process (including the former and the latter) as SSA of electrons in multidimensions. The SSA in multidimensions is different from that discussed in 1D, in that the trapping by the large amplitude waves is no longer important. The new mechanism is more like the ion shock surfing, while in this case the turbulent electrostatic waves play a role of the reflecting wall. We think the self-consistent shock structure in multidimensions, that is, a finite extent of the turbulent region along the shock normal as well as the oblique wavefronts, are important ingredients of the strong electron acceleration.

Let us compare the energy gain estimated from the above argument with the simulation results. The energy gain of electrons from the motional electric field $E$ can be estimated as
\begin{equation}
 \frac{\Delta \epsilon}{1/2 m_e V_0^2} = \frac{e E L}{1/2 m_e V_0^2},
\end{equation}
where $L$ and $V_0$ are the distance the particle travels along the electric field, and the upstream bulk velocity in the downstream rest frame, respectively. Rewriting the electric field by using the relative velocity difference between the background plasma and the particle $V$ as $E = V B / c$, we obtain
\begin{equation}
 \frac{\Delta \epsilon}{1/2 m_e V_0^2} =
  2 \left( \frac{V}{V_0} \right)  \left( \frac{c}{V_0} \right)
  \left( \frac{\wce}{\wpe} \right) \left( \frac{L}{c/\wpe} \right).
\end{equation}

For the estimate of the energy gain within the shock transition region $\epsilon_1$, we use the drift velocity of the reflected ions in the upstream frame $V = V_r$. Since the $x$ and $y$ components of the reflected ions drift velocity measured in the rest frame of the upstream electrons are $V_{r,x}/V_0 \sim -2$ and $V_{r,y}/V_0 \sim 2$, we have $V_r/V_0 = 2 \sqrt{2}$. Substituting the measured penetration distance of the particle $L_1/c/\wpe \sim 5$, we obtain
\begin{equation}
 \frac{\Delta \epsilon_1}{1/2 m_e V_0^2} = 2.8 \times
  \left( \frac{L_1}{c/\wpe} \right)
  \sim 14. \label{estimate1}
\end{equation}
This estimate is smaller than the observed energy gain of $\sim 20$ at $\wpe t \simeq 1110$, suggesting that the particle energy gain arises not only from the motional electric field, but also the wave electric field. The particle is actually accelerated by the large positive $E_x$ at $\wpe t \simeq 1110$ (see \figref{trajectory-exey}). The sum of this additional energy from the wave electric field and that estimated from \eqref{estimate1} agrees well with the observed energy gain. We note that the direct acceleration by the wave electric field should not be expected in periodic simulation models of the shock transition region that have commonly been used in the literature. The difference obviously comes from the assumption of the homogeneity made in the models: A spatial gradient of the wave energy exists in a real shock transition region. Therefore, a possibility for a particle to be accelerated by the wave electric field at the edge of the shock front remains finite.

The second step of the acceleration in the upstream region can also be estimated by assuming $V = V_0$
\begin{equation}
 \frac{\Delta \epsilon_2}{1/2 m_e V_0^2} = \left( \frac{L_2}{c/\wpe} \right)
  \sim 20,
\end{equation}
where a measured distance of $L_2 / c/\wpe \sim 20$ is used. This energy gain is consistent with the simulation result.

We have seen that the energy gains of two acceleration phases are comparable, thereby, both are important for nonthermal particle acceleration. However, we think the former acceleration within the shock transition region plays a more important role. As a result of the first step, energetic electrons are preferentially transported to the negative $x$ direction and eventually reflected back to the upstream region, where they suffer a further energization. Furthermore, the energy gain of the second step is proportional to the Larmor radius of the preaccelerated electron in the upstream, that is determined by the energy gain in the first step.

\section{CONCLUSIONS AND DISCUSSION}
We have studied strong electron acceleration in a high Mach number, perpendicular shock by using a 2D PIC simulation code. We demonstrate that the nonthermal electrons with spectral indices of $2.0 {\rm -} 2.5$ are generated in the shock. The efficient electron energization occurs at the leading edge of the shock transition region through the interactions with large amplitude electrostatic waves produced by the Buneman instability. However, the electrostatic turbulence in 2D has considerably different characteristics compared to that in 1D: The growth of many oblique modes produces multidimensional potential structures. In addition, the wavefronts of the electric fields are oblique to the shock normal and are almost perpendicular to the reflected ion beam. We show that these effects actually play a crucial role for the electron acceleration. The electrostatic turbulence in the shock transition region enhances anomalous transport of energetic electrons toward the upstream and a fraction of electrons are reflected back from the shock front. These reflected electrons suffer a further acceleration by the upstream convective electric field. We call the acceleration mechanism as SSA, however, the new mechanism is more like the classical shock surfing of ions rather than that of electrons discussed previously based on 1D simulations \citep{2001PhRvL..87y5002M,2002ApJ...572..880H}.

It is clear that periodic models of the shock transition region often used to investigate the nonlinear development of beam instabilities are not appropriate to study the electron acceleration process discussed here, because this requires a spatial inhomogeneity inherent in the shock. We have also performed 2D simulations by adopting a periodic simulation model, which is similar to those used in the literature. We find that the electron energization observed in the periodic model is less efficient than that shown in the present paper. Note that our simulation results are basically consistent with those found by \cite{2007ApJ...661L.171O}. Since the spatial inhomogeneity plays an essential role for the electron acceleration mechanism, it is natural that we find the significant differences between numerical simulations of the self-consistent shock and the periodic model. On the other hand, \cite{0741-3335-50-6-065020} modeled perpendicular shocks by colliding two plasma clouds. They found almost planar electrostatic waves, which contradicts the results of our simulations. The use of strong magnetic field ($\wpe/\wce = 5$), or the short simulation time compared to the ion cyclotron period in their simulations might be the reason for this. In short, one should be careful to interpret the results obtained by adopting simplifying assumptions. We find that the particle acceleration in the self-consistent shock structure is actually much more efficient than in the periodic model. We conclude that the SSA can play a role even in multidimensions and will contribute importantly to the nonthermal production at high Mach number shocks, although the mechanism is different from that previously discussed based on 1D PIC simulations. We think that, however, the details of the mechanism are not yet fully understood and several issues remain to be answered.

An interesting question is that ``What is the relative acceleration efficiency between 1D and 2D ?'' We have also performed a 1D simulation with the same parameters. However, we do not find large amplitude electrostatic waves at the leading edge of the foot region; only less intense electrostatic waves, which cause a weaker electron heating, are observed in the deeper shock transition region. Although the condition of the Buneman instability is formally satisfied, the inhomogeneity along the shock normal may prevent the wave growth because the wavelength of the most unstable wave ($\sim 2\pi V_0/\wpe$) is comparable to the convective Larmor radius of electrons ($V_0/\wce$). We observe that the bulk of incoming electrons are merely decelerated at the leading edge of the foot region so as to cancel the current produced by the reflected ions. This observation may suggest that the threshold of the Buneman instability is lower in 2D. The reason for this is that the $y$ component of the relative drift can also contribute to the development of the instability in 2D: Since the system is homogeneous along this direction, the prediction of linear theory will hold. Comparisons with higher Mach numbers and/or weaker magnetic fields, in which the Buneman instability is excited both in 1D and 2D, are needed anyway to discuss the relative efficiency. Nevertheless, if one invokes the 1D simulation results discussed in \cite{2007ApJ...661..190A}, the observed power-law index of the electron energy spectrum was $\sim 3 {\rm - } 4$. The harder spectral index in 2D may suggest that the electron acceleration is even more efficient than 1D. More detailed analysis of the differences between 1D and 2D, as well as the comparisons with the periodic model will be reported elsewhere in the future.

In the context of the electron injection into DSA process, the maximum attainable energy is also important. The efficient electron injection in quasi-perpendicular shock through SSA followed by SDA requires that SSA should accelerate electrons to energies of the order of the upstream bulk ion energy \citep{2007ApJ...661..190A}. Although the present simulation results satisfy the requirement, the mass ratio dependence of the maximum energy is not yet clear, and thus should be investigated in more detail. More specifically, it is easy to expect that the maximum velocity of accelerated electrons depends on the phase velocity of the electrostatic waves, which does not depend on the mass ratio. Hence, one might think that increasing the mass ratio leads to relatively lower maximum energies. However, this argument may not apply when multiple electron reflections occur. In the present study, we have shown the trajectory of an accelerated particle, that is reflected by the shock only once. So far, we do not find any multiply reflected electrons. However, multiple reflections may occur at shocks with different parameters. In the case of the ion shock surfing, multiple reflections are believed to provide an efficient mechanism for injecting low-energy pick-up ions into Fermi acceleration \citep{1996JGR...101..457Z}. We think that the same thing can also happen for electron acceleration. The property of the turbulent region will probably be important for the multiple reflections; namely, the wave amplitude and the width of the turbulent region. \cite{2002ApJ...572..880H} showed that the energy gain by their SSA in 1D is proportional to the amplitude of wave electric field. Since the transport of energetic electrons that plays a key role for the particle acceleration is enhanced by electrostatic waves, the wave amplitude will also be important for SSA in multidimensions. Since the saturation level of the Buneman instability increases with increasing the Mach number, the electron acceleration through multiple reflections may occur at higher Mach number shocks. In addition, we think the width of the turbulent region, in which the Buneman instability provides the dominant electron energization, is also important. The Buneman instability rapidly thermalizes electrons until the temperature approaches the upstream bulk energy, which occurs on an extremely short scale length of the order of $(m_i/m_e)^{1/3} V_0/\wpe$ \citep{1988Ap&SS.144..535P}. Thus, the use of the real mass ratio increases the width of the turbulent region (normalized to the wavelength of the instability) by a factor of $\sim 4$. The dependence of the acceleration efficiency on these quantities will be an another subject of future investigation. Large scale numerical simulation studies as well as theoretical modelings are needed to improve our understanding of the strong electron acceleration process.

In the present study, we consider electron energization by large amplitude electrostatic waves excited by the Buneman instability. However, other instabilities may also contribute to nonadiabatic heating and acceleration of particles in the shock. Since 1D PIC simulations consider only instabilities having wavevectors parallel to the shock normal, the effects of plasma waves propagating in other directions are completely neglected. It is well known that the cross-field current flowing transverse to the magnetosonic shock can be a source of free energy. Such instabilities may play a dominant role at moderate Mach number shocks such as planetary bow shocks in the heliosphere, in which the excitation of the Buneman instability is prohibited due to large electron thermal velocities. It has been pointed out that the enhanced dissipation by microinstabilities can modify the nonstationary behavior of the macroscopic shock structure \citep{2004AnGeo..22.2345S,2005JGRA..11002105S}. We also find some differences in the shock structure between 1D and 2D. For instance, we observe a less nonstationary shock in 2D, and the maximum compressed magnetic field at the overshoot region of $B/B_0 \sim 8$ observed in 2D is significantly reduced from that in 1D $B/B_0 \sim 13$. We think that microinstabilities do play a role of regulating the macroscopic shock structure. In addition, we think the degree of freedom along the magnetic field is also an another important subject. Recently, \cite{2008ApJ...681L..85U} have performed 2D PIC simulations of perpendicular shock, and demonstrated that the electron acceleration efficiency is reduced when the background magnetic field lies in the simulation plane. This is in clear contrast to the present results. We think the reason for this discrepancy is that the oblique wavefronts are produced in our simulations because the reflected ion beam rotates in the plane perpendicular to the magnetic field. Another related work is the 2D simulations of perpendicular shock by \cite{2007GeoRL..3414109H} using a somewhat stronger magnetic field strength, which evidences the emission of oblique whistlers in the shock transition region. According to the authors, these oblique whistlers play a role of suppressing the self-reformation of the shock front. \cite{2006ApJ...653..316B} reported that the larger scale shock surface fluctuations (rippling) enhance the efficiency of SDA in quasi-perpendicular shock. These effects should also be taken into account when one considers realistic electron heating and acceleration efficiencies.

Finally, we would point out that understanding of the injection process is important for the nonlinear coupling between energetic particles and the shock. There are observational indications that the magnetic fields at astrophysical shocks are significantly amplified by orders of magnitude from the typical interstellar value of a few $\mug$ \citep[e.g.,][]{2005ApJ...621..793B,2007Natur.449..576U}. It has been considered that the strong amplification is due to the action by the cosmic rays \citep{2004MNRAS.353..550B}. Although the applicability of the simplified theory is still controversial, the nonlinear feedback by the presence of the cosmic rays will be of great importance. It is indispensable to know the number of injected particles as well as their energy density for understanding of the nonlinear interactions between the shock and energetic particles. The microscopic dynamics in a thin shock transition region will have a nonnegligible impact on the global shock evolution and the efficiency of particle acceleration to cosmic ray energies.

\acknowledgments We thank N.~Shimada, C.~H.~Jaroschek, and K.~Nagata for valuable discussions. T.~A. is indebted to T.~Terasawa, M.~Fujimoto, T.~Yokoyama, M.~Hirahara, and R.~Matsumoto for their helpful suggestions. This work was supported by ISAS/JAXA and the Solar-Terrestrial Environment Laboratory, Nagoya University. T.~A. is supported by a JSPS Research Fellowship for Young Scientists.


\end{document}